\documentclass{eptcs}
\providecommand{\event}{GANDALF 2014} 
\usepackage{hyperref}
\usepackage{breakurl}             

\usepackage{latexsym} 
\usepackage{epic,eepic}
\usepackage{epsfig,wrapfig}

\usepackage[usenames]{color}

\usepackage[pdflatex=true,recompilepics=false]{gastex} 

\usepackage{auto-pst-pdf}
\usepackage{amssymb}
\usepackage{amsthm}
\usepackage{amsmath}

\newcommand{\ignore}[1]{}

\def\buchi{B\"{u}chi }

\def\nat{\mathbb{N}}

\newcommand{\zug}[1]{\langle #1 \rangle}

\newcommand{\calA}{{\mathcal A}}

\newcommand{\EXPTIME}{{\sf EXPTIME}}

\newtheorem{theorem}{Theorem}
\newtheorem{definition}{Definition} 
\newtheorem{lemma}[theorem]{Lemma}
\newtheorem{proposition}[theorem]{Proposition} 

\newcommand{\set}[1]{\{ #1 \} }

\def\pl{\mathit{pl}}
\def\succ{\mathit{succ}}
\def\ctr{\mathit{ctr}}
\def\CARET{{\sc CaRet}}
\def\PATH{{\sc PATH}}
\def\NWTL{{\sc Nwtl}}
\def\LTL{{\sc Ltl}}

\def\call{\mathit{call}}
\def\ret{\mathit{ret}}
\def\En{\mathit{En}}
\def\Ex{\mathit{Ex}}
\def\Calls{\mathit{Calls}}
\def\Retns{\mathit{Retns}}
\def\VPS{{VPA}}
\def\BGame{{$\omega$-MG}}
\def\Bproblem{{$\omega$-MGP}}
\def\VPG{{VPRG}}
\def\MVPG{{MVPG}}
\def\RGG{{RGG}}
\newcommand{\tuple}[1]{\langle #1 \rangle}
\def\internal{\mathit{int}}
\def\push{\mathit{push}}
\def\pop{\mathit{pop}}
\newcommand{\earrow}[1]{\xrightarrow{#1}}

\def\EXPTIME{{\sc Exptime}}
\def\TWOEXPTIME{{\sc 2Exptime}}
\def\PSPACE{{\sc Pspace}}
\def\TM{\mathcal{A}}
\def\obj{\mathit{obj}}

\def\same{\mathit{same}}
\def\up{\mathit{up}}
\def\down{\mathit{down}}

\def\calA{{\cal A}}
\def\calB{{\cal B}}
\def\calP{{\cal P}}

\def\digit{{d}}
\def\pp{\mathcal{C}}
\def\final{\mathit{Fin}}
\def\exits{\mathcal{E}}

\def\ppre{\mathcal{C}_{pre}}
\def\ppost{\mathcal{C}_{post}}

\def\BG{\mathit{CG}}

\begin{document}


\title{
Visibly Pushdown Modular Games 
\thanks{This work was partially funded by the MIUR grants
FARB 2011-2012-2013, Universit\`a degli Studi di Salerno (Italy).} 
}



\author{Ilaria De Crescenzo
\institute{Universit\`a degli Studi di Salerno}
\institute{Dipartimento di Informatica}
\and
Salvatore La Torre
\institute{Universit\`a degli Studi di Salerno}
\institute{Dipartimento di Informatica}
\and
Yaron Velner
\institute{Tel Aviv University}
\institute{The Blavatnik School of Computer Science}
}

\def\titlerunning{Visibly Pushdown Modular Games}
\def\authorrunning{I. De Crescenzo, S. La Torre \& Y. Velner}

\maketitle

%
%

\begin{abstract}

\begin{abstract}
Games on recursive game graphs can be used to reason about the control flow of sequential programs with recursion. In games over recursive game graphs, the most natural notion of strategy is the modular strategy, i.e., a strategy that is local to a module 
and is oblivious to previous module invocations, and thus does not depend on the context of invocation. In this work, we study for the first time modular strategies with respect to winning conditions that can be expressed by a pushdown automaton. 
We show that such games are undecidable in general, and become decidable for visibly pushdown automata specifications. 
Our solution relies on a reduction to modular games with finite-state automata winning conditions, which are known in the literature. 
We carefully characterize the computational complexity of the considered decision problem. In particular, we show that modular games with a universal \buchi or co-\buchi visibly pushdown winning condition are \EXPTIME-complete, and when the winning condition is given by a \CARET\ or \NWTL\ temporal logic formula the problem is \TWOEXPTIME-complete, and it remains \TWOEXPTIME-hard even for simple fragments of these logics.
As a further contribution, we present a different solution for modular games with finite-state automata winning condition that runs faster than known solutions for large specifications and many exits.
\end{abstract}

\end{abstract}


\section{Introduction}

Recursive state machines (RSMs) carefully model the control
flow of systems with potentially recursive procedure calls \cite{AlurRecursiveSMjournal}. A recursive state machine is composed of a set of modules, whose vertices can be standard vertices or can correspond to invocations of other modules.
A large number of hardware and software systems fits into this class, such 
as procedural and object-oriented programs, distributed systems, communication protocols and web services. 

In the open systems setting, i.e., systems where an execution depends on the interaction of the system with the environment, the natural counterpart of recursive state machines 
is two-player recursive game graphs. 
A recursive game graph (\RGG) is  essentially a recursive state machine where vertices are split into two sets each controlled by one of the players, and thus corresponds to pushdown games 
with an emphasis on the modules composing the system.

In this paper we focus on solving pushdown games on \RGG\ in which the first player is restricted to \emph{modular strategies}~\cite{ALM06_TCS}. A strategy is a mapping that specifies, for each play ending into a controlled state, the next move. 
Modular strategies are formed of a set of strategies, one for each   \RGG\ module,
that are \emph{local} to a module and \emph{oblivious} of the history of previous module activations, 
i.e., the next move in such strategies is determined by looking only at the local 
memory of the current module activation (by making the local memory \emph{persistent} across module activations, deciding these games becomes undecidable already with reachability specifications \cite{ALM06_TCS}).

The main motivation for considering modular strategies is related to the synthesis of controllers \cite{PR89,ThomasCav02}:
given a description of the system where some of the choices depend upon the input and some  represent uncontrollable internal non-determinism, the goal is to design a controller that supplies inputs to the system such that 
it satisfies the
correctness specification. Synthesizing a controller thus corresponds to computing
winning strategies in two-player games, and  a modular strategy
would correspond to a modular controller. 

The notion of modular strategy is also of independent interest and has recently found application in other contexts, such as, the automatic transformation of programs for ensuring security policies in privilege-aware operating systems \cite{RepsCav2012}, and  a general formulation of the synthesis problem from libraries of open components \cite{RP2013}. 

The problem of deciding the existence of a modular strategy in a recursive game graph has been already studied with respect to $\omega$-regular 
specifications. The problem is known to be NP-complete for reachability specifications \cite{ALM06_TCS}, \EXPTIME-complete for specifications given as deterministic and universal \buchi or Co-\buchi automata, and  \TWOEXPTIME-complete for \LTL\ specifications \cite{ALM03cav}. 

In this paper, we study this problem with respect to several classes of specifications that can be expressed as a pushdown automaton. 
We show that in the general case
the problem is undecidable. 
We thus focus on visibly pushdown automata  (\VPS) \cite{NWA} specifications
with \buchi or co-\buchi acceptance. 
In the following, we refer to this problem as the \MVPG\ problem and omit the acceptance condition of the \VPS\ by meaning either one of them.

Our main contributions are:
\begin{itemize}
\item
We show a polynomial time reduction from the \MVPG\ problem with deterministic or universal \VPS\ specifications to recursive modular games over $\omega$-regular specifications. By \cite{ALM03cav}, we get that this problem is \EXPTIME-complete. 
We then use this result to show the membership to \TWOEXPTIME\ for the \MVPG\ problem  with nondeterministic \VPS\ specifications.
\item
We show that 
when the winning condition is expressed as a formula of the temporal logics  \CARET \\ \noindent~\cite{caret} and \NWTL ~\cite{nwtl} the \MVPG\ problem 
is \TWOEXPTIME-complete, and hardness can be shown also for very simple fragments
of the logics. 
In particular, we show a \TWOEXPTIME\ lower bound for the fragment containing only conjunctions of disjunctions of bounded-size path formulas (i.e., formulas expressing either the requirement that a given finite sequence is a subsequence of a word or its negation), 
that is in contrast with the situation in finite 
game graphs where \PSPACE-completeness holds for larger significant fragments  (see~\cite{ltlgeneratorsJournal,concur03}). 
On the positive side, we are able to show an exponential-time algorithm to decide the \MVPG\ problem for specifications given as conjunctions of  temporal logic formulas that can be translated into a polynomial-size \VPS\ (such formulas include the path formulas). 
\item
We also give a different solution for recursive games with finite-state automata specifications.
Our approach yields an upper bound of $|G|\,2^{O(d^2 (k+\log d)+\beta)}$ for the \MVPG\ problem, where $d$ is the number of $P$ (the \VPS) states, $k$ is the number of $G$ (the \RGG) exits, and $\beta$ is the number of \emph{call edges} of $G$, i.e., the number of module pairs $(m,m')$ such that there is  
a call from $m$ to $m'$.
The known solution~\cite{ALM03cav} yields an $|G|\,2^{O(k d^2 \log(kd))}$ upper bound.
Thus, our solution is faster when $k$ and $d$ are large, and matches the known \EXPTIME\ lower bound~\cite{ALM03cav}.
In addition we use one-way nondeterministic/universal tree automata instead of two-way alternating tree automata, thus we explicitly handle aspects that are hidden in the construction from \cite{ALM03cav}. 
\end{itemize}

\vspace*{-6pt}
\paragraph{Related work.}
Besides the already mentioned work that has dealt with modular games, but only 
for  $\omega$-regular specifications~\cite{ALM03cav} or reachability \cite{ALM06_TCS}, 
other research on pushdown games have focused on the standard notion of winning 
strategy. 
We recall that determining the existence of a standard winning strategy (i.e., non-modular) in pushdown games with reachability specifications is known to be \EXPTIME-complete \cite{Wal01}. Deciding such games is \TWOEXPTIME-complete for nondeterministic visibly pushdown specifications and 3\EXPTIME-complete for \LTL\ and \CARET\ specifications \cite{vpg}. 

The synthesis from recursive-component libraries defines a pushdown game which is \emph{orthogonal} to the \MVPG\ problem: there the modules are already synthesized and the game is on the function calls. Deciding such games for \NWTL\ is \TWOEXPTIME-complete \cite{VardiGandalf}. 
The synthesis from open recursive-component libraries combines both this synthesis problem and the \MVPG\ synthesis. 
Deciding the related game problem with reachability specifications is \EXPTIME-complete \cite{RP2013}.
%
%
Other synthesis  problems dealing with compositions of component libraries are 
\cite{VardiFossacs,AMM2014}, and for modules expressed as terms of the $\lambda Y$-calculus, \cite{SW2013}.

 \vspace*{-7pt}
\section{Preliminaries}\label{sec:prel}
%

\vspace*{-3pt}
Given two positive integers $i$ and $j$, $i\leq j$, we denote with
$[i,j]$ the set of integers $k$ with $i\leq k \leq j$, and
with $[j]$ the set $[1,j]$. 

We fix a set of atomic propositions $AP$ and a finite alphabet $\Sigma$.
A $\omega$-\emph{word} over $\Sigma$ is a mapping that assigns to each \emph{position} $i\in\mathbb{N}$ a symbol $\sigma_i\in\Sigma$, and is denoted as $\{\sigma_i\}_{i\in \mathbb{N}}$ or equivalently $\sigma_1\sigma_2\ldots$.

\vspace*{-8pt} \paragraph{ Recursive game graph.}
A recursive game graph (\RGG) is composed of \emph{game modules} that are essentially two-player graphs (i.e., graphs whose vertices are partitioned into two sets depending on the player who controls the outgoing moves)  with \emph{entry} and \emph{exit} nodes and two different kind of vertices: the nodes and the boxes.  
A \emph{node} is a standard graph vertex and a \emph{box} corresponds to invocations of other game modules in a potentially recursive manner (in particular, entering into a box corresponds to a module \emph{call} and exiting from a box corresponds to a \emph{return} from a module).

\begin{wrapfigure}[10]{r}{4.0truecm}
\scriptsize
%
%

\vspace*{-1.3truecm}
\begin{center}

\unitlength=1.25pt

\begin{gpicture}(0,0)(0,0)
\put(-63,-50){\framebox(90,95){}}

\node[Nw=70,Nh=40,Nmr=3](A)(-16.5,20){} 

\nodelabel[ExtNL=y,NLangle=151,NLdist=0](A){$M_{in}$}

\node[Nw=70,Nh=40,Nmr=3](B)(-16.5,-24){} 

\nodelabel[ExtNL=y,NLangle=210,NLdist=1.5](B){$M_{1}$}



\node[fillcolor=black, Nw=2,Nh=5,Nmr=0](Call1)(-29.5,20){}

\node[Nw=10,Nh=10](Ein)(-44,20){$e_{in}$}

\node[Nw=20,Nh=13,Nmr=2](M1)(-20,20){$b\,$:$\,M_1$} 

\node[fillcolor=black, Nw=3,Nh=3](Exit1)(-10.5,20){} 

\node[Nw=10,Nh=10,Nmr=0](S1)(10,30){$u_1$} 

\nodelabel[ExtNL=y,NLangle=90,NLdist=1](S1){$p_c$}

\node[Nw=10,Nh=10](S2)(10,10){$u_2$} 

\nodelabel[ExtNL=y,NLangle=90,NLdist=1](S2){$p_d$}

\node[Nw=1,Nh=1,linewidth=0, linecolor=white](void)(-57,20){}

\drawedge(Ein,Call1){}

\drawedge(Exit1,S1){}

\drawedge(Exit1,S2){}

\drawedge(void,Ein){}

\drawbpedge[ELpos=30,ELside=r](S1,45,10,Call1,-30,-30){}

\drawbpedge[ELpos=30,ELside=r](S2,265,10,Call1,30,-30){}


\node[Nw=10,Nh=10,Nmr=0](E1)(-44,-24){$e_1$}

\node[Nw=10,Nh=10](S3)(-17.5,-10){$u_3$} 

\nodelabel[ExtNL=y,NLangle=270,NLdist=1](S3){$p_a$}

\node[Nw=10,Nh=10](S4)(-17.5,-38){$u_4$} 

\nodelabel[ExtNL=y,NLangle=90,NLdist=1](S4){$p_b$}

\node[Nw=10,Nh=6, Nmr=0](EX1)(10,-24){$ex_1$}

\drawedge(E1,S3){}

\drawedge(E1,S4){}

\drawedge(S3,EX1){}

\drawedge(S4,EX1){}

\node[Nw=1,Nh=1,linewidth=0, linecolor=white](void1)(-57,-24){}

\node[Nw=1,Nh=1,linewidth=0, linecolor=white](void2)(25,-24){}

\drawedge(void1,E1){}

\drawedge(EX1,void2){}

\end{gpicture}

\end{center}
\vspace*{-0.6truecm}
\caption{A sample \RGG.}\label{fig:graph}

\end{wrapfigure}

As an example consider the  \RGG\ in Fig.~\ref{fig:graph}, where the vertices of player $0$ ($\pl_0$) are denoted with rounds, those of player $1$ ($\pl_1$) with squares and the rectangles denote the vertices where there are no moves that can be taken by any of the players and correspond to calls and exits.
Atomic propositions $p_a$, $p_b$, $p_c$ and $p_d$ label the vertices. 
Each \RGG\ has a distinct game module which is called the \emph{main} module (module $M_{\mathit{in}}$ in the figure). In analogy to many programming languages, we require that the main module cannot be invoked by any other module. 
A typical play starts in vertex $e_{in}$. From this node, there is only one possible move to take and thus the play continues at the call to $M_1$ on box $b$, which then takes the play to the entry $e_1$ in $M_1$. This is a vertex of the adversary, who gets to pick the transition and thus can decide to visit either $u_3$ (generating $p_a$) or $u_4$ (generating $p_b$). In any of the cases, the play will evolve reaching the exit and then the control will return to module $M_1$ at the return vertex on box $b$. 
 Here $\pl_0$ gets to choose if generating $p_c$ or $p_d$ and so on back to the call to $M_1$. 
 Essentially, along any play alternatively  $\pl_1$ chooses one between $p_a$ and $p_b$, and $\pl_0$ chooses one between $p_c$ and $p_d$.
Formally, we have the following definitions.
\begin{definition}({\sc Recursive Game Graph})
A \emph{recursive game graph} $G$ over $AP$ is a triple $(M, m_{in},$ $\{S_m\}_{m \in M} )$ where $M$ is a finite set of module names, $m_{in} \in M$ denotes the main module and for each $\ m \in M$, $ S_m$ is a game module.
A \emph{game module} $S_m$ is 
$(N_m, B_m, Y_m,\En_m, \Ex_m, \delta_m, \eta_m, P^0_m, P^1_m)$ where:
\begin{itemize}
\item $N_m$ is a finite set of nodes and $ B_m$ is a finite set of boxes;
\item $Y_m : B_m \rightarrow (M\setminus \{m_\mathit{in}\})$ maps every box to a module;
\item $\En_m\subseteq N_m$  is a non-empty 
set of entry nodes;
\item 
$\Ex_m\subseteq N_m$  is a (possibly empty)
set of exit nodes; 
\item $\delta_m : N_m \cup\Retns_m \rightarrow 2^{N_m \cup \Calls_m}$ is a transition function where $\Calls_m=\{(b,e)| b \in B_m , e \in \En_{Y_m(b)}\}$ is the set of calls and  $\Retns_m=\{(b,e)| b \in B_m , e \in \Ex_{Y_m(b)}\}$ is the set of returns; 
\item $\eta_m : V_m \rightarrow 2^{AP}$ labels in $2^{AP}$
each vertex from $V_m = N_m \cup \Calls_m \cup \Retns_m$;

\item $ P^0_m$ and $P^1_m$ form a partition of $(N_m \cup \Retns_m)\setminus \Ex_m$; $P^0_m$ is the set of the positions of 
 $\pl_0$ and  $P^1_m$ is the set of the positions of  $\pl_1$.
\end{itemize} 
\end{definition}
In the rest of the paper, we denote with: 
$G$ an \RGG\ as in the above definition;  
$V=\bigcup_m V_m$ (set of vertices); 
$B=\bigcup_m B_m$ (set of boxes); 
$\Calls=\bigcup_m \Calls_m$ (set of calls); 
$\Retns=\bigcup_m \Retns_m$ (set of returns); 
$\Ex=\bigcup_m \Ex_m$ (set of exits); 
$P^\ell= \bigcup_m P^\ell_m$ for $\ell\in[0,1]$ (set of all positions of $\pl_\ell$);
and $\eta: V \rightarrow  2^{AP}$ such that $\eta(v)=\eta_m(v)$ where $v\in V_m$.



\noindent
To ease the presentation we make the following assumptions (with $m\in M$):
\vspace*{-4pt}
\begin{itemize}
\item there is only one entry point to every module $S_m$ and we refer to it as $e_m$;


\item  there are no transitions to an entry, i.e.,  $e_m\not\in\delta_m(u)$ for every $u$;

\item there are no transitions from an exit, i.e.,  $\delta_m(x)$ is empty for every $x \in \Ex_m$;

\item a module is not called immediately after a return from another module, i.e., 
$\delta_m(v) \subseteq N_m$ 
for every $v\in \Retns_m$. 

\end{itemize} 
\vspace*{-4pt}

A (global) \emph{state} of an  \RGG\ is composed of a call stack and a vertex, that is, each state of $G$ is 
of the form $(\alpha,u)\in B^* \times V$ where $\alpha=b_1\ldots b_h$,  
$b_1\in B_{m_{in}}$, $b_{i+1}\in B_{Y(b_i)}$ for $i\in[h-1]$ and  $u\in V_{Y(b_h)}$. 

A \emph{play} of $G$ is 
 a (possibly finite) 
 sequence of states $s_0 s_1  s_2\ldots$ such that $s_0=(\epsilon, e_\mathit{in})$ and for $i\in \mathbb{N}$,
denoting $s_i=(\alpha_i, u_i)$, one of the following holds:\\
\hspace*{0.28truecm}$-$ {\bfseries Internal move:} $u_i \in (N_m \cup \Retns_m)\setminus  \Ex_m$, 
$u_{i+1} \in \delta_m(u_i)$ and $\alpha_i = \alpha_{i+1}$;\\
\hspace*{0.28truecm}$-$ {\bfseries Call to a module:} $u_i \in \Calls_m$, $u_i=(b, e_{m'})$,  $u_{i+1}=e_{m'}$ and $\alpha_{i+1}=\alpha_i.b$;\\
\hspace*{0.28truecm}$-$ %
{\bfseries Return from a call:} $u_i \in \Ex_m$, $\alpha_i=\alpha_{i+1}.b$, and $u_{i+1}=(b,u_i)$.

Fix an infinite play $\pi=s_0 s_1\ldots$ of $G$ where $s_i=(\alpha_i,u_i)$ for each $i\in\nat$. 

With $\pi_k$ we denote $s_0\ldots s_k$, i.e., the prefix of $\pi$ up to $s_k$. 
%
%
For a finite play $\pi'.s$, with $\ctr(\pi'.s)$ we denote the module $m$ where the control is at $s$, i.e., 
such that $u\in V_m$ where $s=(\alpha,u)$.
We define $\mu_\pi$ such that $\mu_\pi(i,j)$ holds iff for some $m\in M$, 
$u_i\in \Calls_{m}$ and 
$j$ is the smallest index s.t. $i<j$,
$u_j\in \Retns_{m}$ and $\alpha_i=\alpha_j$ ($\mu_\pi$ captures the matching pairs of calls and returns in $\pi$).

%

%


\paragraph{Modular strategies.}
Fix  $\ell\in[0,1]$. 
A \emph{strategy} of $\pl_\ell$ is a function $f$ that associates a legal move to every play ending in a node controlled by $\pl_\ell$. 

A modular strategy constrains the notion of strategy by allowing only to define the legal moves depending on the \lq\lq local memory\rq\rq of a module activation (every time a module is re-entered the local memory is reset). 
 %
%

Formally, a \emph{modular strategy} $f$ of $\pl_\ell$ is a set of functions $\{ f_m \}_{m \in M}$, one for each module $m\in M$, where $f_m : V^*_m.P^\ell_m \rightarrow V_m$ is such that $f_m(\pi.u)\in \delta_m(u)$ for every $\pi \in V^*_m, u \in P^\ell_m$.

The local successor 
of a position in $\pi$ 
is: the successor according to the matching relation $\mu_\pi$ at matched calls,  
undefined   
at an exit or an unmatched call, 
and the next position otherwise. 
Formally, the \emph{local successor} of $j$, denoted  $\succ_\pi(j)$,  is:
$h$ if $\mu_\pi(j,h)$ holds;  otherwise,
is undefined if either $u_{j}\in \Ex$ 
or $u_j\in \Calls$ and $ \mu_\pi(j,h)$ does not hold for every $h>j$;
and $j+1$ in all the remaining cases.

For each $i\le |\pi|$, the \emph{local memory} of $\pi_i$, denoted $\lambda(\pi_i)$, is  the maximal sequence $u_{j_1}\ldots u_{j_k}$ such that $u_{j_k}=u_i$ and $j_{h+1}=\succ_\pi(j_h)$ for each $h\in[k-1]$. (Note that since the sequence is maximal, $u_{j_1}=e_m$ where $m=\ctr(\pi_i)$.)

A play $\pi$ \emph{conforms to} a modular strategy $f=\{ f_m \}_{m \in M}$ of $\pl_\ell$ 
if for every $i < \mid\pi\mid $, denoting $\ctr(\pi_i)=m$,   $u_i\in P^\ell_m$ implies that
$u_{i+1}=f_m(\lambda(\pi_i))$. 

 Consider again the example from Fig.~\ref{fig:graph}. A strategy of $\pl_0$ that chooses alternatively to generate $p_c$ and $p_d$ is modular, in fact it requires as memory just to store the last move from the return of $b$, and thus 
 is local to the current (sole) activation of module $M_{in}$. Instead, a strategy that attempts to match each $p_a$ with $p_c$ and each $p_b$ with $p_d$ is clearly non modular. 
 
 
We remark that modular strategies are oblivious to the previous activations of a module. 
In the \RGG\ of Fig.~\ref{fig:graph}, a modular strategy for $\pl_1$ would only allow either one of the behaviors: \lq\lq $\pl_1$ always picks $p_a$\rq\rq or \lq\lq $\pl_1$  always picks $p_b$\rq\rq.  


\vspace*{-10pt} 

\paragraph{Winning conditions and modular games.}
A modular game on RGG is a pair $\langle G, L\rangle$ where $G$ is an RGG and $L$ is a winning condition.
A winning condition is a set $L$ of $\omega$ words over a finite alphabet $\Sigma=2^{AP}$, where $AP$ is a set of propositions.
Given an \RGG\ $G$, for a play $\pi=s_0s_1\ldots$ of $G$, with $s_i=(\alpha_i,u_i)$, we define the word $w_\pi=\eta(u_0)\eta(u_1)\ldots$, which is the mapping that assigns to each position the corrispondent symbol from $\Sigma$.
A (modular) strategy $f$ is winning if $w_\pi \in L$ for every play $\pi$ of $G$ that conforms to $f$.
The \emph{modular} \emph{game problem} asks to determine the existence 
 of a winning (modular) strategy of $\pl_0$ in a given modular game. 
In the following sections, we consider $L$ given by pushdown, visibly pushdown automata and by \LTL\ , \CARET\ or \NWTL\ formulas.

\ignore{
Given an \RGG\ $G$ consider a language $L$. 

For a play $\pi=s_0s_1\ldots$ of $G$, with $s_i=(\alpha_i,u_i)$, we define the word $w_\pi$ as $\sigma_0\sigma_1\ldots$ such that for $i\in\nat$, $\sigma_i=(\mu_m(s_i),t_i)$ where
$\ctr(\pi_i)=m$ and $t_i$ is $\call$ if $u_i\in \Calls$, $\ret$ if $u_i\in \Retns$, and 
$\internal$ otherwise.

A (modular) strategy $f$ is winning for $\langle G,P\rangle$  if $w_\pi$ is accepted by $P$  
for every play $\pi$ that conforms to $f$.
}

\vspace*{-7pt} 
\section{Pushdown specification}
\vspace*{-3pt} 
%
 
\paragraph{Pushdown modular games.}
A \emph{pushdown modular game} is a pair $\langle G,\calP \rangle$ where $G$ is an RGG and $\calP$ is a pushdown automaton, whose accepted language defines the winning condition in $G$.
A \emph{pushdown automaton} $\calP$ is a tuple $(Q,q_0,\Sigma, \Gamma,
\delta, \gamma^\bot, F)$ where $Q$ is a finite set of states, $q_0 \in Q$ is the initial state, $\Sigma$ is a finite alphabet, 
$\Gamma$ is a finite stack alphabet, $\gamma^\bot$ is the
bottom-of-stack symbol, $F \subseteq Q$ defines an acceptance condition, and 
$\delta: Q \times \{\Sigma\cup\epsilon \} \times \Gamma\rightarrow 2^{Q \times \Gamma^*}$ is the transition function. A pushdown automaton is \emph{deterministic} if it satisfies the following two conditions: -$\delta(q,\alpha,\gamma) $ has at most one element for any $q \in Q$, $\gamma \in \Gamma$ and $\alpha\in\Sigma$ (or $\alpha=\epsilon$); - if $\delta(q,\epsilon,\gamma)\neq\emptyset$ for any $q\in Q$ and $\gamma\in \Gamma$ then $\delta(q,\alpha,\gamma)=\emptyset$ for any $\alpha\in\Sigma$.

\vspace*{-10pt}
\paragraph{Undecidability of pushdown specification.} 
The modular game problem becomes undecidable if we consider winning conditions expressed as standard (deterministic) pushdown automata. 
This is mainly due to the fact that the stack in the specification pushdown automaton 
is not synchronized with the call-return structure of the recursive game graph.

We prove the undecidability of our problem with pushdown specification by presenting a reduction from the problem of checking the emptiness of the intersection of two deterministic context-free languages.

Consider two context-free languages $L_1$ and $L_2$  on an alphabet $\Sigma=\{\sigma_1, \sigma_2,...,\sigma_n\}$, which are accepted by two pushdown automata, $\calP_1$ and $\calP_2$, respectively. We want to construct an instance $\tuple{G,\calP}$ of a deterministic pushdown modular game problem such that exists a winning modular strategy for $\pl_0$ in $\tuple{G,\calP}$ if and only if the intersection of $L_1$ and $L_2$ is not empty.

\begin{wrapfigure}[16]{r}{6cm}
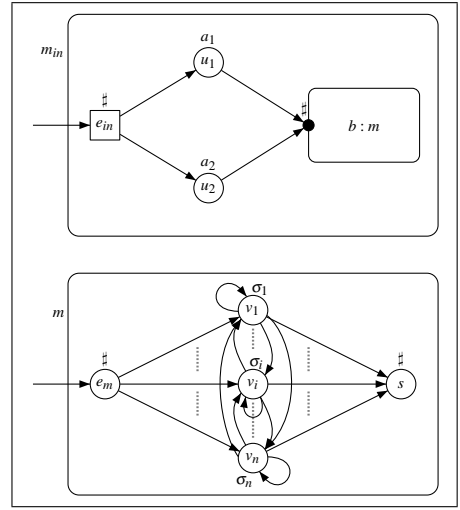

\tiny

\vspace*{-0.7truecm}
\begin{center}

\unitlength=1.4pt

\begin{gpicture}(50,10)(-30,21)
\put(-65,-103){\framebox(120,136){}}

 \node[Nw=100,Nh=60,Nmr=3](MIN)(0,0){} 
 \nodelabel[ExtNL=y,NLangle=160,NLdist=1](MIN){$m_{in}$}

\node[Nw=8,Nh=8,Nmr=0](e1)(-40,0){$e_{in}$}    
\nodelabel[ExtNL=y,NLangle=90,NLdist=1](e1){$\sharp$}
\node[Nw=8,Nh=8](a2)(-12,-17){$u_{2}$}
\nodelabel[ExtNL=y,NLangle=90,NLdist=1](a2){$a_2$}
\node[Nw=8,Nh=8](a1)(-12,17){$u_{1}$}
\nodelabel[ExtNL=y,NLangle=90,NLdist=1](a1){$a_1$}

\node[Nw=30,Nh=20,Nmr=2](box)(30,0){$b : m$} 
\node[fillcolor=black, Nw=3,Nh=3](Call)(15,0){}
\nodelabel[ExtNL=y,NLangle=105,NLdist=1](Call){$\sharp$}

\node[Nw=1,Nh=1,linewidth=0, linecolor=white](void)(-60,0){}

\drawedge(e1,a1){}
\drawedge(e1,a2){}
\drawedge(a1,Call){}
\drawedge(a2,Call){}
\drawedge(void,e1){}

 \node[Nw=100,Nh=60,Nmr=3](M)(0,-70){} 
 \nodelabel[ExtNL=y,NLangle=160,NLdist=1](M){$m$}

\node[Nw=1,Nh=1,linewidth=0, linecolor=white](voidm)(-60,-70){}
\node[Nw=8,Nh=8](em)(-40,-70){$e_{m}$}    
\nodelabel[ExtNL=y,NLangle=90,NLdist=1](em){$\sharp$}
\node[Nw=8,Nh=8](fin)(40,-70){$s$} 
\nodelabel[ExtNL=y,NLangle=90,NLdist=1](fin){$\sharp$}

\node[Nw=8,Nh=8](s1)(0,-50){$v_1$} 
\nodelabel[ExtNL=y,NLangle=70,NLdist=0](s1){$\sigma_1$}
\node[Nw=8,Nh=8](si)(0,-70){$v_i$} 
\nodelabel[ExtNL=y,NLangle=80,NLdist=0](si){$\sigma_i$}
\node[Nw=8,Nh=8](sn)(0,-90){$v_n$} 
\nodelabel[ExtNL=y,NLangle=-110,NLdist=1](sn){$\sigma_n$}

\drawline[AHnb=0,linewidth=0.6,linecolor=gray, dash={0.5}{0.5}](0,-55)  (0,-61)
\drawline[AHnb=0,linewidth=0.6,linecolor=gray, dash={0.5}{0.5}](0,-75)  (0,-85)

\drawline[AHnb=0,linewidth=0.6,linecolor=gray, dash={0.5}{0.5}](-15,-60)  (-15,-67)
\drawline[AHnb=0,linewidth=0.6,linecolor=gray, dash={0.5}{0.5}](-15,-72)  (-15,-79)

\drawline[AHnb=0,linewidth=0.6,linecolor=gray, dash={0.5}{0.5}](15,-60)  (15,-67)
\drawline[AHnb=0,linewidth=0.6,linecolor=gray, dash={0.5}{0.5}](15,-72)  (15,-79)

\drawloop[loopdiam=7, loopangle=150](s1){}
\drawloop[loopdiam=5, loopangle=-90](si){}
\drawloop[loopdiam=7, loopangle=-30](sn){}

\drawbpedge[ELpos=30,ELside=r](s1,-60,10,si,30,10){}
\drawbpedge[ELpos=30,ELside=r](s1,0,10,sn,30,20){}

\drawbpedge[ELpos=30,ELside=r](sn,-60,-10,si,210,10){}
\drawbpedge[ELpos=30,ELside=r](sn,240,10,s1,210,20){}

\drawbpedge[ELpos=30,ELside=r](si,-60,-10,s1,210,10){}
\drawbpedge[ELpos=30,ELside=r](si,-60,10,sn,30,10){}

\drawedge(em,s1){}
\drawedge(em,si){}
\drawedge(em,sn){}

\drawedge(s1,fin){}
\drawedge(si,fin){}
\drawedge(sn,fin){}

\drawedge(voidm,em){}

\end{gpicture}

\end{center}
\vspace*{-0.7truecm}
\caption{The module $m_{in}$ and $m$}\label{fig:mainReduction}

\end{wrapfigure}

The basic idea of the reduction is to construct a game where
 $\pl_1$ challenges $\pl_0$ to generate a word from either $L_1$ or 
$L_2$, and $\pl_0$ must match the choice of $\pl_1$ without knowing it in order 
to win.   
We construct an \RGG\ $G$ with two modules, $m_{in}$ and $m$ (see the Fig.~\ref{fig:mainReduction}). 

The module $m_{in}$ is the main module and  is composed of an entry $e_{in}$, two internal nodes $u_1$ and $u_2$, and one box $b$ labeled with $m$. The entry $e_{in}$ belongs to $\pl_1$ and has two transitions, one to each internal node. From $u_1$ and $u_2$ there is only one possible move, which leads to $b$. The labeling function associates the symbol $a_1$ to $u_1$ and the symbol $a_2$ to $u_2$ with $a_1,a_2\notin \Sigma$. The node $e_{in}$ and the call $(b,e_{m})$ are both labeled with $\sharp \notin \Sigma \cup \set{a_1,a_2}$. Observe that 
since the only choice of $\pl_1$ is at $e_{in}$, for any strategy $f$ of $\pl_0$ there are only two plays conforming to it: 
one going through $u_1$ and the other through $u_2$.

The module $m$ is essentially a deterministic generator of any word in $\set{\sharp}.\Sigma^*.\set{\sharp}$. The module $m$ has one entry $e_m$, 
$|\Sigma|$ internal nodes $v_1, v_2, ..., v_n$ and a sink node $s$
(i.e., a node with only ingoing edges). 
All the vertices of $m$ belong to $\pl_0$. There are only outgoing edges from $e_m$, which take to each of the other vertices of $m$. 
Moreover, there is a transition from $v_i$ to $v_j$ for any ${i,j} \in [n]$, and from any node there is a move to $s$. Each node $v_i$ is labeled with  $\sigma_i$. 
for $i\in [n]$. The symbol $\sharp$ labels $e_m$ and $s$. 

As winning condition, we construct a deterministic pushdown automaton $\calP$. 
In the inital state $\calP$ reads $\sharp$ and moves into state $q_0$.
Fix $i\in[2]$. From $q_0$ and on input $a_i$, $\calP$ enters state $q_i$. 
From $q_i$, $\calP$ reads two occurrences of $\sharp$ and enters 
the initial state of $\calP_i$.  
From any $\calP_i$ state, $\calP$ behaves as $\calP_i$ 
and in addition, 
from each final state of $\calP_i$, it has a move on input $\sharp$  
that takes to the only final state $q_f$.

Since the strategy must be modular, in the module $m$ the player $\pl_0$ has no information about the choice  of $\pl_1$ in $m_{in}$. Also, the local strategy 
in module $m$ generates one specific word (there are no moves of $\pl_1$ allowed 
in $m$) and thus this is the same independently of the moves of $\pl_1$ 
in module $m_{in}$. Thus, the local strategy in $m$ is winning if and only if
it generates a word in the intersection of $L_1$ and $L_2$, and therefore, the following theorem holds.
\begin{theorem}\label{thrm:undecidable}
The (deterministic) pushdown modular game problem is undecidable.
\end{theorem}


\section{Solving modular games with \VPS\ specifications} \label{sec:VPAreduction}
\paragraph{Visibly pushdown automata.} 
Consider the finite alphabet $\Sigma$, and let $\call$, $\ret$, and $\internal$ be new symbols.
We denote with $\Sigma_\call=\Sigma\times\set{\call}$, 
$\Sigma_\ret=\Sigma\times\set{\ret}$ and
$\Sigma_\internal=\Sigma\times\set{\internal}$ and with $\widehat{\Sigma}=\Sigma_\call \cup \Sigma_\ret \cup \Sigma_\internal$. 

%
A 
\emph{visibly pushdown automaton} (\VPS) $P$ is a tuple $(Q,Q_0,\widehat{\Sigma}, \Gamma\cup\set{\gamma^\bot},
\delta, F)$ where $Q$ is a finite set of states, $Q_0 \subseteq Q$ is a set
of initial states, $\widehat{\Sigma}$ is a finite alphabet, 
$\Gamma$ is a finite stack alphabet,  $\gamma^\bot$ is the
bottom-of-stack symbol, $F \subseteq Q$ defines an acceptance condition, and $\delta=\delta^\internal\cup\delta^\push \cup \delta^\pop$ where $\delta^\internal\subseteq Q \times \Sigma_\internal \times Q$, 
$\delta^\push\subseteq 
Q \times \Sigma_\call \times 
 \Gamma \times Q$, and $\delta^\pop\subseteq Q \times \Sigma_\ret \times (\Gamma\cup\{\gamma^\bot\})
\times Q$.

 A \emph{configuration} (or global state) 
 of $P$ is a pair $(\alpha,q)$ where $\alpha\in\Gamma^*.\{\gamma^\bot\}$ and $q\in Q$. 
 Moreover, $(\alpha,q)$ is \emph{initial} if $q \in Q_0$ and  $\alpha=\gamma^\bot$.
We omit 
the semantics of the transitions of $P$ being quite standard. It can be obtained similarly to that of \RGG\ with the addition of the inputs. 
Here we just observe that we allow pop transitions on empty stack (a stack containing only the symbol $\gamma^\bot$). In particular, a pop transition do not change the stack when $\gamma^\bot$ is at the top, and  by the definition of $\delta^\push$, $\gamma^\bot$ cannot be pushed onto the stack.
A \emph{run} $\rho$ of $P$ over the input $\sigma_0\sigma_1\ldots$ is an infinite sequence 
$C_0\earrow{\sigma_0}C_{1}\earrow{\sigma_1} \ldots $ where $C_0$ is the initial configuration and such that,  
for each $i\in\nat$, $C_{i+1}$ is obtained from $C_i$ by applying a transition on input $\sigma_i$.


Acceptance of an infinite run depends on the control states that are visited infinitely often. 
Fix a run $\rho=(\gamma^\bot,q_0)\earrow{\sigma_0}(\alpha_1,q_1)\earrow{\sigma_1} (\alpha_2,q_2)\ldots $.
With a  \emph{B\"uchi acceptance condition},
$\rho$ is accepting if $q_i\in F$ for infinitely many $i\in \nat$ (\emph{B\"uchi} \VPS). 
With a  \emph{co-B\"uchi  acceptance condition}, 
$\rho$ is accepting if there is a $j\in \nat$ such that $q_i\not\in F$ for all $i>j$
(\emph{co-B\"uchi} \VPS). 



A \VPS\  $P$ is \emph{deterministic} if:  (1) $|Q_0|=1$,
(2) for each $q\in Q$ and $\sigma\in \Sigma_\call\cup\Sigma_\internal$
there is at most one transition of $\delta$ from $q$ on input $\sigma$, 
and (3) for each $q_1\in Q$, $\sigma\in \Sigma_\ret$, $\gamma\in \Gamma\cup \set{\gamma^\bot}$ there is at most one transition from $q$ on input $\sigma$ and stack
symbol $\gamma$.
%
Note that a deterministic \VPS\ is such that for each word $w$ there is at most a run  over it. 

 \noindent 
For a word $w$, a \emph{deterministic/nondeterministic} (resp. \emph{universal}) 
\VPS\ accepts $w$ if there exists an accepting run 
over $w$ (resp. all runs over $w$ are accepting). 
\vspace*{-6pt}
\paragraph{Visibly pushdown games.}
A \emph{visibly pushdown game on \RGG} (\VPG) is a pair 
$\langle G,P\rangle$ where $G$ is an \RGG\ and $P$ is a visibly pushdown automaton (see \cite{NWA}). Consider a \VPG\ $\langle G,P\rangle$ where $G$ is an \RGG\ and $P$ is a \VPS . 
For a play $\pi=s_0s_1\ldots$ of $G$, with $s_i=(\alpha_i,u_i)$, we define the word $w_\pi$ as $\sigma_0\sigma_1\ldots$ such that for $i\in\nat$, $\sigma_i=(\eta_m(s_i),t_i)$ where
$\ctr(\pi_i)=m$ and $t_i$ is $\call$ if $u_i\in \Calls$, $\ret$ if $u_i\in \Retns$, and 
$\internal$ otherwise.
The \emph{visibly pushdown} (\emph{modular}) \emph{game problem} asks to determine the existence 
 of a (winning) modular strategy of $\pl_0$ in a given \VPG\ such that $w_\pi$ is accepted by $P$  
for every play $\pi$ that conforms to $f$.
We denote the visibly pushdown modular game problem as  \MVPG\ problem.

When the VPA is a finite state automaton $\calB=(Q, q_0, \Sigma,\delta, F)$ we denote with 
\Bproblem\ the \emph{$\omega$-modular game problem} that asks to determine the existence 
 of a winning modular strategy of $\pl_0$ in a given $\langle G,\calB \rangle$ game (\BGame).

\paragraph{Solving games with \VPS\ specifications} \label{par:Reduction}
We consider games with winning conditions that are given by a VPA with different acceptance conditions.
We present a reduction from recursive games with VPA specifications to recursive games with specifications that are given as finite state automata. 
The reduction is almost independent of the acceptance condition, and it works for reachability and safety conditions as well as for \buchi and co-\buchi acceptance conditions.

\begin{wrapfigure}[12]{r}{6cm}
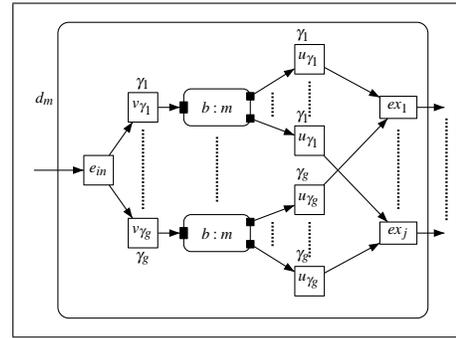

\tiny
%
%

\vspace*{-1.0truecm}
\begin{center}

\unitlength=1.4pt

\begin{gpicture}(50,150)(-30,-20)
\put(-65,-45){\framebox(120,90){}}

 \node[Nw=100,Nh=80,Nmr=3](MIN)(-3,0){} 
 \nodelabel[ExtNL=y,NLangle=160,NLdist=1](MIN){$d_{m}$}

\node[Nw=8,Nh=8,Nmr=0](e1)(-42,0){$e_{in}$}    
\node[Nw=8,Nh=8,Nmr=0](a2)(-30,-17){$v_{\gamma_{g}}$}
\nodelabel[ExtNL=y,NLangle=-90,NLdist=1](a2){$\gamma_{g}$}
\node[Nw=8,Nh=8,Nmr=0](a1)(-30,17){$v_{\gamma_{1}}$}
\nodelabel[ExtNL=y,NLangle=90,NLdist=1](a1){$\gamma_{1}$}

\node[Nw=8,Nh=8,Nmr=0](g2a)(15,30){$u_{\gamma_{1}}$}
\nodelabel[ExtNL=y,NLangle=105,NLdist=1](g2a){$\gamma_{1}$}
\node[Nw=8,Nh=8,Nmr=0](g2b)(15,8){$u_{\gamma_{1}}$}
\nodelabel[ExtNL=y,NLangle=105,NLdist=1](g2b){$\gamma_{1}$}   
\node[Nw=8,Nh=8,Nmr=0](g1a)(15,-8){$u_{\gamma_{g}}$}   
\nodelabel[ExtNL=y,NLangle=105,NLdist=1](g1a){$\gamma_{g}$}   
\node[Nw=8,Nh=8,Nmr=0](g1b)(15,-30){$u_{\gamma_{g}}$}   
\nodelabel[ExtNL=y,NLangle=105,NLdist=1](g1b){$\gamma_{g}$}   
\node[Nw=10,Nh=6, Nmr=0](ex1)(39,17){$ex_{1}$}   
\node[Nw=10,Nh=6, Nmr=0](ex2)(39,-17){$ex_{j}$}   

\node[Nw=18,Nh=10,Nmr=2](box1)(-10,-17){$b : m$} 
\node[Nw=18,Nh=10,Nmr=2](box2)(-10,17){$b : m$} 
\node[fillcolor=black, Nw=2,Nh=3, Nmr=0](Call1)(-19,-17){}
\node[fillcolor=black, Nw=2,Nh=3, Nmr=0](Call2)(-19,17){}

\node[Nw=1,Nh=1,linewidth=0, linecolor=white](void)(-60,0){}
\node[Nw=1,Nh=1,linewidth=0, linecolor=white](void1)(52,17){}
\node[Nw=1,Nh=1,linewidth=0, linecolor=white](void2)(52,-17){}

\node[fillcolor=black, Nw=2,Nh=2, Nmr=0](rtn1)(-1,20){}
\node[fillcolor=black, Nw=2,Nh=2, Nmr=0](rtn2)(-1,14){}
\node[fillcolor=black, Nw=2,Nh=2, Nmr=0](rtn3)(-1,-14){}
\node[fillcolor=black, Nw=2,Nh=2, Nmr=0](rtn4)(-1,-20){}

\drawline[AHnb=0,linewidth=0.6,linecolor=black, dash={0.5}{0.5}](-30,11)  (-30,-11)
\drawline[AHnb=0,linewidth=0.6,linecolor=black, dash={0.5}{0.5}](-10,9)  (-10,-9)
\drawline[AHnb=0,linewidth=0.6,linecolor=black, dash={0.5}{0.5}](15,24)  (15,14)
\drawline[AHnb=0,linewidth=0.6,linecolor=black, dash={0.5}{0.5}](15,-14)  (15,-24)

\drawline[AHnb=0,linewidth=0.6,linecolor=black, dash={0.5}{0.5}](5,21)  (5,14)
\drawline[AHnb=0,linewidth=0.6,linecolor=black, dash={0.5}{0.5}](5,-14)  (5,-21)

\drawline[AHnb=0,linewidth=0.6,linecolor=black, dash={0.5}{0.5}](39,11)  (39,-11)

\drawline[AHnb=0,linewidth=0.6,linecolor=black, dash={0.5}{0.5}](52,14)  (52,-14)

\drawedge(e1,a1){}
\drawedge[ELside=r](e1,a2){}
\drawedge(a1,Call2){}
\drawedge(a2,Call1){}
\drawedge(void,e1){}

\drawedge(rtn1,g2a){}
\drawedge(rtn2,g2b){}
\drawedge(rtn3,g1a){}
\drawedge(rtn4,g1b){}

\drawedge(g1b,ex2){}
\drawedge[ELside=r, ELpos=60](g2b,ex2){}
\drawedge[ELpos=60](g1a,ex1){}
\drawedge(g2a,ex1){}

\drawedge(ex1,void1){}
\drawedge(ex2,void2){}

\end{gpicture}

\end{center}
\vspace*{-0.5truecm}
\caption{The module $d_{m}$. All the vertices of $d_m$ are controlled by $\pl_1$}\label{fig:ReductionYaron}

\end{wrapfigure}
The reduction transforms a recursive game graph with a visibly pushdown automaton specification (with some acceptance condition)
to a slightly different recursive game graph with a finite-state automaton specification (with the same type of acceptance condition).
The key idea is to embed the top stack symbol of a VPA $P$ in the states of a finite-state automaton $A$. In addition, the states of $A$ will simulate the corresponding states of $P$ and thus we will get that the winning conditions are equivalent.
Clearly, a finite-state automaton cannot simulate an unbounded stack.
While it is easy to keep track of the top symbol after a $\push$ operation, extracting the top symbol after a $\pop$ operation requires infinite memory.
For this purpose we exploit the fact that the stacks of the VPA $P$ and the game graph $G$ are synchronized and we introduce a new \emph{dummy} module $d_m$ for every module in $G$.
Recall that the invocation of a module $m$ in $G$ correspond to a $\push$ operation in $P$.
We replace every invocation of $m$ by a call to $d_m$.
In $d_m$ (see Figure~\ref{fig:ReductionYaron}) $\pl_1$ first has to \emph{declare} the value of the top symbol in $P$ before the $\push$ operation by going to the corresponding $v_{\gamma_1},\dots,v_{\gamma_g}$ state in $d_m$, and $A$ can verify that $\pl_1$ is honest since it keeps track of the current top-symbol (if the player is not honest then $A$ goes to a sink accepting state and $\pl_1$ loses).
After the declaration, the module invokes the actual module $m$ and when $m$ terminates, then $\pl_1$ must declare again the top-symbol $\gamma_i$ of $P$, visiting the vertex $u_{\gamma_i}$,
and $A$ changes his simulated top symbol accordingly.

Denoting with $k$ the number of exits of the starting \RGG, with $g$ the number of stack symbols, and $d$ the number of states of the specification, we get that 
the resulting \RGG\ $G$ has $2k$ exits and the resulting automaton $A$ has $O(dg)$ states. Thus combining this with the solution from \cite{ALM03cav}, we get an upper bound linear in $|G|$ and exponential in $2k (d\,g)^2 \log(2\,k\,d\,g)$. We have:

\begin{theorem}\label{theo:det}
The  \MVPG\ problem with winning conditions expressed as either a deterministic B\"uchi \VPS\ or a deterministic co-B\"uchi \VPS\ is \EXPTIME-complete.
\end{theorem}
The proposed reduction can be extend for universal \VPS\ specification. W.l.o.g we assume that in the non-deterministic \VPS\  for every state, stack letter and labeling there are exactly two possible transitions. In this case we add a dummy module $e$, that is composed only by $\pl_1$ nodes and has one exit. Each transition from a node $v$ to a node $u$ is splitted in two transitions, $v\rightarrow e$ and $e\rightarrow u$. In the module $e$, $\pl_1$ resolves the nondeterminism, selecting one of the two possible transitions for the \VPS\ specification.
The choices of $\pl_1$ in $e$ are oblivious to $\pl_0$.
Hence, the universal \VPS\ accepts if and only if $\pl_0$ has a strategy that wins against all $\pl_1$ choices in $e$, and we get the next theorem.
\begin{theorem}\label{theo:uni}
The  \MVPG\ problem with winning conditions expressed as a universal B\"uchi  (resp. co-B\"uchi) \VPS\  is \EXPTIME-complete.
\end{theorem}
We can handle nondeterministic VPAs in the following way:
Let $P$ be a nondeterministic B\"uchi \VPS\ $P$.
By \cite{NWA}, we can construct a nondeterministic B\"uchi \VPS\ $P'$ that 
accepts a word $w$ iff $P$ does not accept it, and such that the size of $P'$ is exponential in the size of $P$. 
Complete $P'$ with transitions that take to a rejecting state such that for each word there is at least a run of $P'$ over it.
Let $P''$ be the dual automaton of $P'$, i.e., $P$ has the same components of $P'$ except that acceptance is now universal and the set of accepting states is now interpreted as a  co-B\"uchi condition. Clearly, $P''$ accepts exactly the same words as $P$ and has size exponential in $|P|$. 
Similarly, we can repeat the above reasoning starting from a co-B\"uchi \VPS\ $P$.
 Therefore, we have:
\begin{theorem}\label{theo:nondet}
The  \MVPG\ problem with winning conditions expressed as a nondeterministic B\"uchi  (resp. co-B\"uchi) \VPS\  is in \TWOEXPTIME.
\end{theorem}

\section{Temporal logic winning conditions}
By \cite{caret}, we know that given a \CARET\ formula $\varphi$ it is possible to 
construct a nondeterministic B\"uchi \VPS\ of size exponential in $|\varphi|$ 
that accepts exactly all the words that satisfy $\varphi$. From \cite{nwtl}, 
we know that the same holds for the temporal logic \NWTL. 
Thus, given a formula $\varphi$ in any of the two logics, 
we construct a  B\"uchi \VPS\ $P$ for its negation $\neg \varphi$. By dualizing as in the case of nondeterministic \VPS\ specifications, 
we get a co-B\"uchi \VPS\ that accepts all the models of $\varphi$ and which size 
is exponential in $|\varphi|$. Since both \CARET\ and \NWTL\ subsume \LTL\ \cite{pnueli77}, and \LTL\ games are known to be \TWOEXPTIME-hard~\cite{PR89}  already on standard finite game graphs, we get:
\begin{theorem}\label{theo:TL}
The  \MVPG\ problem with winning conditions expressed as \CARET\ and \NWTL\ formulas is  \TWOEXPTIME-complete.
\end{theorem}

 The complexity of the  temporal logic  \MVPG\ problem remains 2\EXPTIME -hard even if we consider simple
 fragments. 

A \emph{path formula} is a formula expressing either the requirement that a given sequence appears as a subsequence in an $\omega$-word or its logical negation. 
Path formulas are captured by \LTL\ formulas of the form
$\Diamond(p_1\wedge \Diamond(p2\wedge\ldots\Diamond(p_{n-1}\wedge \Diamond p_n) \ldots ))$ and by their logical negation,  
 where each $p_i$ is state predicate, $\Diamond \psi$ (\emph{eventually} $\psi$) denotes  that $\psi$ holds at some future position, and $\wedge$ is the Boolean conjunction. We denote such a fragment of \LTL\ as \PATH-\LTL.

We present a reduction from exponential-space alternating Turing machines.
We only give here the general idea.
We use a standard encoding of computations, where cell contents are preceded by the cell number written in binary 
($2N$ atomic propositions suffice to encode $2^N$ cell numbers) 
and configurations are sequences of cells encodings  ended with a marker
(the tape head and the current control state are encoded as cell contents).

Denote with $Q$ and $\Sigma$ respectively the control states and
the input alphabet, and let $2^N$ be the number of cells used in each configuration.
A configuration encoding  is a sequence of the form $\zug{0}\sigma_0\ldots\zug{2^N}\sigma_{2^N}$ where there is a $i$ s.t. $\sigma_i\in Q\times \Sigma_i$ (this denotes the current state, the symbol of cell $i$ and that the tape head is on cell $i$), $\sigma_j\in \Sigma$ for all $j\neq i$ (symbol in cell $j$), and 
$\zug{h}$ is the binary encoding of $h$ (cell number) over new symbols 
$d^\top_r$ and $d^\bot_r$ for $r\in [N]$ ($d^\top_r$ is equivalent to $1$ and $d^\bot_r$ to $0$ in the binary encoding). 
A path of a computation (computations of alternating TM can be seen as trees of configurations) is encoded as a sequence $C_0d_0\$\ldots C_id_i\$   \ldots$ where each $C_i$ is a configuration ($C_0$ is initial)
and $d_i$ is the transition taken from $C_i$ to $C_{i+1}$.

We construct an \RGG\ $G$ with two modules $M_{\mathit{in}}$ and $M_{1}$. 
In $M_{\mathit{in}}$, initially, $\pl_0$ generates an encoding of an initial configuration,  then, a transition is selected by $\pl_0$, if the initial state is existential, or  by $\pl_1$, 
otherwise. In both cases, an end-of-configuration marker $\$$ is
generated and then $\pl_0$ is in charge to generate again a configuration encoding, 
and so on. A call to $M_1$ is placed before generating the first cell encoding of each configuration and after generating each cell encoding.
In $M_1$, $\pl_1$ selects one among a series of actions that can either 
state that everything is fine ($\mathit{ok}$) or that some check is required (by raising one over nine objections). $M_1$ has only one exit.

The goal of $\pl_0$ is to build an encoding of an accepting run of a TM $\TM$ on a given input word, while the goal of $\pl_1$ is to point out errors in such encoding by raising objections to delimit the cell encodings where the check has to take place. There are two possible mistakes that can occur: the $i^\mathit{th}$ cells of two consecutive configurations do not conform the transition relation of $\TM$ and the  number of a cell is not the successor of the number of the preceding cell in the configuration. 
We use separate groups of objections to point out each of these mistakes.
These specifications can be captured with a formula $\varphi$ defined as 
$\{\neg\Diamond\mathit{obj} \vee [(\psi_{\mathit{wr1}}\vee \psi_{\Delta}) \wedge (\psi_{\mathit{wr2}}\vee\psi_\sharp)]\} \wedge
(\Diamond \mathit{obj} \vee \Diamond F)$ where: 
(1) $\mathit{obj}$ denotes that an objection has been raised; 
(2) $F$ denotes a state predicate  that is true on symbols of the encoding that correspond to final states of $\TM$; (3)  $\psi_{\mathit{wr1}}$  and $\psi_{\mathit{wr2}}$ capture all the illegal uses of respectively the first and the second type of objections;  (4) $\psi_{\Delta}$ checks the transition relation between two consecutive configurations on the cells selected by the raised objections of type $1$; and (5) 
$\psi_\sharp$ checks the correct encoding of the cell numbers of two consecutive cells selected by the raised objections of type $2$.
All the above formulas can be written with disjunctions of path formulas except for $\psi_\sharp$ that is a conjunction of disjunctions of path formulas. 
Using De Morgan laws, the total formula can be transformed into an equivalent formula 
of  size polynomial in $|\varphi|$,
which is a disjunction of conjunctions of path formulas. 
Also note that all the used path formulas are of bounded size (the most complex one uses eleven occurrences of $\Diamond$). 
Moreover, in a modular strategy $\pl_0$ cannot  use the fact that $\pl_1$ has raised an objection to decide the next move since the objections are raised in a different module (which has just one exit).  Therefore, in order to win, $\pl_0$ must correctly generate the computations of the TM.  
We get the following: 
\begin{lemma}\label{lem:2exptime}
The  \MVPG\ problem with winning conditions expressed 
as a conjunction of disjunctions of bounded-size \PATH-\LTL\ formulas is \TWOEXPTIME-hard.
\end{lemma}

It is known that each formula $\varphi$ from  
\PATH-\LTL\ admits a deterministic B\"uchi word automaton accepting all the models of $\varphi$ and which is 
linear in its size \cite{ltlgeneratorsJournal}. The same can be shown with B\"uchi \VPS, 
by extending \PATH-\LTL\ allowing the versions of the $\Diamond$ operator of \CARET\ and \NWTL. 
By the closure properties of universal visibly pushdown automata we can easily extend Theorem~\ref{theo:uni} to winning conditions given as intersection of deterministic \VPS s and thus: 
\begin{theorem}\label{theo:frag}
The  \MVPG\ problem with winning conditions expressed as a conjunction of \CARET\ and \NWTL\ formulas that admit a deterministic B\"uchi or co-B\"uchi 
\VPS\ generator of polynomial size is \EXPTIME-complete.
\end{theorem}

Now consider the larger fragment of formulas $\bigvee_{i=1}^h\bigwedge_{j=1^k} \varphi_{i,j}$ where for each $\varphi_{i,j}$ we can construct either a deterministic B\"uchi or a deterministic co-B\"uchi \VPS\ $P_{i,j}$ of polynomial size 
that generates all the models of $\varphi_{i,j}$.  We are only able to show 
an \EXPTIME\ lower bound using a construction similar to
that used in the reduction of Lemma~\ref{lem:2exptime} 
 However, we observe that a matching upper bound cannot be shown using automata constructions, since we would need to
 manage the union of $N$ specifications without an exponential blow-up, and since intersection is easy, this would contradict  Lemma~\ref{lem:2exptime}.

\section{Improving the tree automata construction to solve \Bproblem\ with B\"uchi condition}\label{sec:gameToTree}


We assume the standard definitions of trees and  nondeterministic/universal 
tree automata with B\"uchi and co-B\"uchi acceptance (universality refers to the fact that all runs must be accepting in order to accept. See \cite{ALM03cav} for definitions). 

\vspace*{-8pt}
\paragraph{\bf General structure of the construction.}
 Fix a \BGame\ $\langle G,\calB \rangle$ where 
$\calB=(Q, q_0,\Sigma,\delta,  F)$ is a 
deterministic B\"uchi automaton and $G$ is as in Section \ref{sec:prel}. 
We construct a B\"uchi tree automaton ${\calA}_{G,\calB}$ that accepts a tree if and only if $\pl_0$ has a winning modular strategy in the game $\zug{G,\calB}$.

%
%
The trees accepted by $\calA_{G,\calB}$ must encode $G$ and a modular strategy on it
(\emph{strategy trees}). Each such tree essentially has
a subtree rooted at a child of the
root for each module of $G$ and each such subtree is the unwinding of the corresponding module along with a labeling encoding the strategy.
 
The general idea is to check on each subtree of the root some properties 
of the corresponding local function of the encoded modular strategy, 
by assuming some other properties on the local 
functions of the other modules (as in an assume-guarantee reasoning). 
These assumptions concern: 
a call structure $\BG$ (\emph{B\"uchi call graph}), to handle acceptance on plays involving infinitely many unreturned calls;
a set $\exits$ for each module, each giving a superset of the exits that can be visited during the plays;
a set of extended pre-post conditions $\pp$, that for each module $m$, each 
exit $x$ and each possible state $q$ of $\calB$, carries the requirement that if $m$ is entered with $q$ and the play exits at $x$, then this must happen 
with state $q'$ such that $(m,q,x,q')\in \pp$
and after visiting an accepting state if this is required by $\pp$. 
To ensure correctness all modules must share the same assumptions, 
thus $\calA_{G,\calB}$ guesses the assumptions at the root and then passes them 
onto the children of the root. 

The tasks of ${\calA}_{G,\calB}$ thus are: 
recognizing the strategy trees; 
ensuring the correctness of the extended pre-post conditions; 
ensuring
that all the plays according to the modular strategy conform the B\"uchi call graph and do not exit a module from an exit different from those listed in $\exits$;
checking the acceptance condition of $\calB$ on all the plays encoded in the strategy tree, using the pre-post condition, the acceptance condition of $\calB$ and the B\"uchi call graph.

The above tasks are split among the 
B\"uchi tree automata  ${\calA}_{G}$, ${\calA}^{\exits}_\BG$ and ${\calA}_{\calB,\pp,\BG}$. 
The automata ${\calA}_{G}$ and ${\calA}^{\exits}_\BG$ are nondeterministic and 
check the fulfillment of the properties related to the game graph. ${\calA}_{G}$ is in charge of 
verifying that the input tree is a valid strategy tree.
${\calA}^{\exits}_\BG$ is parameterized over the set of exits $\exits$ and a B\"uchi call graph $\BG$. 
%
%
The automaton ${\calA}_{\calB,\pp,\BG}$, which is a universal B\"uchi tree automaton, checks the extended pre-post condition $\pp$, 
simulates $\calB$ and checks the fulfillment of its winning conditions.

${\calA}_{G,\calB}$ captures the intersection of ${\calA}_{G}$ and the automaton that, 
at the root, nondeterministically guesses $\exits, \BG, \pp$, and 
then at the children of the root, captures the intersection of  ${\calA}_{\calB,\pp,\BG}$ and 
${\calA}^{\exits}_\BG$. 


\vspace*{-8pt} 
\paragraph{\bf The automaton ${\calA}_{G}$.}
Let $k$ be the maximum over the number of exits of $G$ modules and the out-degree of 
$G$ vertices. Denote with  $\Omega_G$ the set $\set{\mathit{dummy},\mathit{root}}\cup (V\setminus P^0) \cup(P^0\times [k])$ (recall $V$ denotes the set of vertices of $G$).
${\calA}_{G}$ accepts strategy trees, i.e., $\Omega_G$-labeled $k$-trees
that encode modular strategies of $\pl_0$. 
%
%
%

\begin{wrapfigure}[10]{r}{5.4cm}
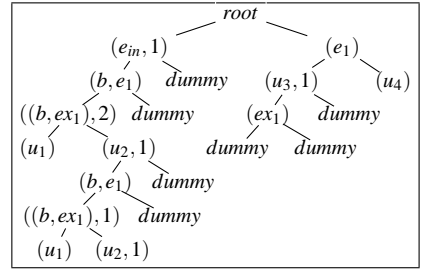

\scriptsize
\setlength{\unitlength}{0.45mm}
\vspace*{-0.4truecm}
\begin{center}
\begin{gpicture}(105,40)(-28,-28)
\put(-37.5,-35){\framebox(120,78){}}

\node[Nw=15,Nh=20, linecolor=white](root)(30,40){$root$}
\node[Nw=26,Nh=8, linecolor=white](Ein)(0,30){$(e_{in},1)$}
\node[Nw=26,Nh=7,linecolor=white](E1)(60,30){$(e_{1})$}
\drawedge[AHnb=0](root,Ein){}
\drawedge[AHnb=0](root,E1){}
\node[Nw=26,Nh=7,linecolor=white](Call1)(-7,20){$(b, e_1)$}
\node[Nw=12,Nh=6,linecolor=white](dummy)(17,20){$dummy$}
\node[Nw=20,Nh=8,linecolor=white](U3)(45,20){$(u_{3},1)$}

\node[Nw=10,Nh=8,linewidth=0,linecolor=white](U4)(75,20){$(u_{4})$}

\drawedge[AHnb=0](Ein,Call1){}

\drawedge[AHnb=0](Ein,dummy){}

\drawedge[AHnb=0](E1,U3){}

\drawedge[AHnb=0](E1,U4){}

\node[Nw=12,Nh=6,linewidth=0,linecolor=white](EX1)(38,10){$(ex_{1})$}
\node[Nw=12,Nh=6,linecolor=white](dummy4)(62,10){$dummy$}

\node[Nw=20,Nh=6,linewidth=0,linecolor=white](Ret1)(-21,10){$((b,ex_1), 2)$}
\node[Nw=12,Nh=6,linecolor=white](dummy2)(7,10){$dummy$}

\drawedge[AHnb=0](Call1,Ret1){}
\drawedge[AHnb=0](Call1,dummy2){}
\drawedge[AHnb=0](U3,EX1){}
\drawedge[AHnb=0](U3,dummy4){}


\node[Nw=12,Nh=7,linewidth=0,linecolor=white](U1)(-30,0){$(u_1)$}

\node[Nw=22,Nh=7,linewidth=0,linecolor=white](U2)(-3,0){$(u_2, 1)$}

\drawedge[AHnb=0](Ret1,U1){}

\drawedge[AHnb=0](Ret1,U2){}

\node[Nw=12,Nh=6,linewidth=0,linecolor=white](Call1b)(-10,-10){$(b,e_1)$}
\node[Nw=12,Nh=6,linecolor=white](dummy3)(13,-10){$dummy$}

\drawedge[AHnb=0](U2,Call1b){}
\drawedge[AHnb=0](U2,dummy3){}

\node[Nw=20,Nh=7,linewidth=0,linecolor=white](Ret1b)(-20,-20){$((b,ex_1),1)$}
\node[Nw=12,Nh=6,linecolor=white](dummy5)(9,-20){$dummy$}
\drawedge[AHnb=0](Call1b,Ret1b){}
\drawedge[AHnb=0](Call1b,dummy5){}

\node[Nw=20,Nh=8,linewidth=0,linecolor=white](U1b)(-25,-30){$(u_1)$}

\node[Nw=20,Nh=8,linewidth=0,linecolor=white](U2b)(-5,-30){$(u_2,1)$}

\drawedge[AHnb=0](Ret1b,U1b){}

\drawedge[AHnb=0](Ret1b,U2b){}

\node[Nw=12,Nh=6,linecolor=white](dummy6)(29,0){$dummy$}
\node[Nw=12,Nh=6,linecolor=white](dummy7)(55,0){$dummy$}
\drawedge[AHnb=0](EX1,dummy6){}
\drawedge[AHnb=0](EX1,dummy7){}
\end{gpicture}

\end{center}
\vspace*{-0.7truecm}
\caption{A fragment of a strategy tree.}\label{fig:strategyTree}
\end{wrapfigure}

Intuitively, in a strategy tree, 
the label $\mathit{root}$ is associated with the root of the tree. The children of the root are labeled with the entries of each module in $G$. 
A subtree rooted in one of these vertices corresponds to the unrolling of a module. 
If a vertex is labeled with a node that belongs to $pl_0$, the move according to the encoded strategy is annotated with the index of the selected successor. If a node is associated to a call, then
its children are labeled with the matching returns. The $\mathit{dummy}$ nodes are used to complete the $k$-tree.  
A similar formal definition is given \cite{ALM03cav}.

In Fig.~\ref{fig:strategyTree} we depict the top fragment of a strategy tree for $\pl_0$ of the \RGG\ from Fig.~\ref{fig:graph}
\noindent
Given a tree $T$, the automaton ${\calA}_{G}$ accepts T iff it is a strategy tree for G. A construction for ${\calA}_{G}$ can be easily obtained from $G$, and thus we omit it (see \cite{ALM03cav} for a similar construction).  

\begin{proposition}\label{prop:strat}
There exists an effectively constructible B\"uchi 
tree automaton  of size $O(|G|)$ that accepts a $\Omega_G$-labeled k-tree if and only if it is a strategy tree.
\end{proposition}

\vspace*{-4pt}
Directly from the definitions, the following holds:
\vspace*{-4pt}
\begin{proposition}\label{prop:treeStrat}
For a \BGame\ and fixed a player $\pl$, there exists a one-to-one mapping between the modular strategies of $\pl$ and the strategy trees.
\end{proposition}

\vspace*{-10pt} 
\paragraph{\bf The automaton ${\calA}^{\exits}_\BG$.} 

 Fix a modular strategy $f$ of $\pl_0$ in $G$.  
 A \emph{call graph} (of $G$) according to $f$ is a directed graph $(V, \rightarrow)$, $V\subseteq M$,  such that for each play $\pi$ which conforms to $f$,
if a module $m \in M$ is reachable on $\pi$ then $m\in V$ and 
if a call from  $m'$ to $m''$ is done on $\pi$ 
then $m' \rightarrow m''$ holds.

A \emph{ B\"uchi call graphs} $\BG$ of $G$ according to $f$ and $\calB$ is $(V, \rightarrow, \rightarrow_F)$ where $V\subseteq M\times Q$,  $\rightarrow_F \subseteq \rightarrow$ and: (1) denoting with $\xi((m,q))=m$, the graph defined by all the edges $\xi(v)\rightarrow'\xi(v')$ s. t. $v\rightarrow v'$ is a call graph of $G$ according to $f$ (denoted $\xi(\BG)$ in the following), and
(2) for each cycle $v_1 \rightarrow v_2 \rightarrow\ldots\rightarrow  v_{h}\rightarrow  v_{{h+1}}$  (with $v_{1}=v_{{h+1}}$), there exists at least a $j \in [1,h]$ such that $(v_{j}\rightarrow_{F} v_{{j+1}})$ (in order to fulfill a B\"uchi condition, in a cycle of calls there must exist at least a module $m_{i_j}$ among $m_{i_1},\ldots, m_{i_h}$ in which an $F$ state of $\calB$ is visited infinitely often).

In a strategy tree, a node is \emph{enabled} if it is the root or the corresponding vertex of $G$ is reachable in the encoded strategy.  
For a given B\"uchi call graph $\BG$ and a selection of exits $\exits$, 
by assuming that the input tree is a strategy tree, 
the automaton ${\calA}^{\exits}_\BG$ checks that indeed the input tree conforms 
to $\BG$ and $\exits$. 
 
\begin{lemma}\label{lemma:exits}
There exists an effectively constructible B\"uchi tree automaton ${\calA}^{\exits}_\BG$ such that if $ T$ is a strategy tree of $G$ and $f$ is the 
corresponding modular strategy, then ${\calA}^{\exits}_\BG$ accepts 
$T$ unless either 
(1) an exit not in $\exits$ is enabled, or 
(2) $\xi(\BG)$ is not a call graph of $G$ according to $f$ (i.e., there exists an enabled call from $m$ to $m'$ but no edge from $m$ to $m'$ in $\xi(\BG)$). 
The size of ${\calA}^{\exits}_\BG$ is linear in $|G|$. 
\end{lemma}

\vspace*{-10pt} 
\paragraph{\bf The automaton ${\calA}_{\calB,\pp,\BG}$.}

The automaton          
${\calA}_{\calB,\pp,\BG}$ is parameterized over the B\"uchi automaton $\calB$, 
an extended pre-post condition $\pp$ (which essentially summarizes the  effects of $\calB$ executions in each module of $G$) and a B\"uchi call-graph $\BG$.
It is quite complex and its tasks are: 
\vspace*{-2pt}
\begin{enumerate}
\item to simulate $\calB$ on a strategy tree (and in this, it uses $\pp$ s.t. when simulating $\calB$ in a module it is not needed to follow the calls to other modules and the simulation can continue at a matching return);
\vspace*{-2pt}
\item to check the correctness of the pre-post condition $\pp$;
\vspace*{-2pt}
\item to check that the accepting states of $\calB$ are entered
consistently with $\BG$ on the cycles of calls;
\vspace*{-2pt}
\item to check the fulfillment of $\calB$ acceptance conditions. 
\end{enumerate}
\vspace*{-2pt}
We first construct an automaton ${\calA}_{\calB}^\pp$ which ensures task $1$, then on the top of ${\calA}_{\calB}^\pp$ we construct three different automata $\calA_{\calB,\pp}$,$\calA_{\calB,\BG}$ and $\calA_{\calB_{win}}$,  one for each of the  remaining three tasks. 
We then get ${\calA}_{\calB,\pp,\BG}$ by taking the usual cross product for the intersection of these automata (note that an efficient construction can be obtained by discarding all the states that do not agree on the ${\calA}_{\calB}^\pp$ part, thus avoiding a cubic blow-up in the size of ${\calA}_{\calB}^\pp$).  
Under the assumption that the input tree is a strategy tree and $\xi(\BG)$ is consistent with it,  we get that 
${\calA}_{\calB,\pp,\BG}$ accepts only winning strategy trees (i.e., strategy trees that correspond to winning modular strategies) that conform to $\BG$ and $\pp$.
The details on all the above automata are given in the rest of this section. 
Thus, we get:
\begin{lemma}\label{lemma:winning}
Let $\langle G,\calB\rangle$ be a \BGame\ and $\calB$ be a B\"uchi automaton. 
Given a B\"uchi call graph $\BG$ and a pre-post condition $\pp$, there exists an effectively constructible universal B\"uchi tree automaton 
${\calA}_{\calB,\pp,\BG}$ s.t.: if $T$ is a strategy tree of $G$, $f$ is the 
corresponding modular strategy and $\xi(\BG)$ is a call graph of $G$ according to $f$, then ${\calA}_{\calB,\pp,\BG}$ accepts 
$T$ iff $T$ is winning in $\langle G,\calB\rangle$ and consistent with 
$\BG$ and $\pp$.  
The size of ${\calA}_{\calB,\pp,\BG}$  is quadratic in the number of $\calB$ states. 
\end{lemma}

\paragraph{Winning strategy trees.}
For a strategy tree $T$ of $G$, a \emph{play of $T$} is an $\omega$-sequence of $T$-nodes 
$x_1 x_2\ldots$ such that $x_1$ is the child of the root corresponding to $m_{in}$, 
$\alpha_1=\varepsilon$ (\emph{call-stack}) and
for $i\in\nat$, $x_i$ is an enabled node and:
(1) if $x_i$ is labeled with a call to $m$, then $x_{i+1}$ is the child of the 
$T$ root  corresponding to module $m$ (and thus is labeled with the entry $e_m$), and
$\alpha_{i+1}=\alpha_i.x_i$;
(2) if $x_i$ is labeled with an exit $ex$, then $\alpha_{i}=\alpha_{i+1}.y$, $y$ is a node labeled with a call $(b,e_m)$ and
$x_{i+1}$ is the child of $y$ labeled with the return $(b,ex)$;
(3) otherwise, 
$x_{i+1}$ is an enabled child of $x_i$ and $\alpha_{i+1}=\alpha_i$.

Note that any play of a strategy tree $T$ corresponds to a play of $G$ (conforming to the modular strategy defined by $T$). 
 %
A \emph{winning strategy tree} $T$ w.r.t. $\calB$ is such that for all the plays $\nu$ of $T$, $w_\nu$ is accepted by $\calB$. From Proposition \ref{prop:treeStrat}, we get:
\vspace*{-4pt}
\begin{lemma}\label{lemma:winningST}
Given a \BGame\ $\zug{G,\calB}$, a modular strategy is winning iff the corresponding strategy tree is winning.   
\end{lemma}

\vspace*{-10pt} 
\paragraph{Pre-post conditions.}

A \emph{pre-post condition} on the graph $G$ is a pair $\zug{\ppre,\ppost}$ where $\ppre \subseteq M\times Q$ (set of \emph{pre-conditions}), $\ppost\subseteq M\times Q \times \Ex \times Q$ (set of \emph{pre-post conditions}), and such that for each $(m, q, ex_j, q') \in \ppost$, also $(m, q) \in \ppre$ (i.e., tuples of $\ppost$ add a post-condition to some of the pre-conditions of $\ppre$). 


Intuitively, a pre-post condition is meant to summarize all the $\calB$ locations that can be reached on entering each module of $G$ along any play of $T$, and for each reachable exit $ex$, 
all the pairs of $\calB$ locations $(q,q')$ s.t. there exists a play of $T$ along with $\calB$ enters a
module at $q$ and exits it from $ex$ at $q'$. 

Fix a pre-post condition $\zug{\ppre,\ppost}$.

\noindent
 $\zug{\ppre,\ppost}$ is \emph{consistent} with a strategy tree $T$ if for each play $\nu=x_1 x_2\ldots$ of $T$ and for each $x_i$ which is labeled with a call to module $m\in M$, denoting with 
$q$ the location at which the only run of  $\calB$ over $w_{\nu_i}$ ends:
(1) $(m,q)$ belongs to $\ppre$ and (2) if $\nu$ reaches the matching return at $x_{j+1}$, $x_j$ is labeled with exit $ex$ and the location at which the run of $\calB$ over $w_{\nu_{j_1}}$
is $q'$ (i.e., the location when reading the symbol of $ex$ at $x_j$), then $(m,q,ex,q')$ belongs to $\ppost$.       
 


$\zug{\ppre,\ppost}$ 
is \emph{consistent} with a set of exits $\zug{\exits^m}_{m\in M}$ iff:
(1) $\forall m \in M$, $\forall ex\in \exits^m$, if there is a $(m,q)\in \ppre$ 
then there is at least a tuple of the form $(m, q, ex, q')\in \ppost$; 
(2) $\forall (m, q, ex, q')\in \ppost$, then $ex \in \exits^m$ holds.

We extend pre-post conditions with a function $\final: \ppost \rightarrow \{ true, false\} $. 
An \emph{extended pre-post condition} $\pp=\zug{\ppre,\ppost,\final}$ is \emph{consistent} with a strategy tree $T$ if $\zug{\ppre,\ppost}$ is consistent with $T$ and for each play $\nu=x_1 x_2\ldots$ of $T$ s.t. $x_i$ is labeled with a call to module $m\in M$, $x_{j+1}$ is labeled with its matching return, and the portion of run of $\calB$ from $i$ to $j$ starts at location $q$ and ends at location $q'$: 
whenever $\final(m, q, ex, q') = true$ then a location in $F$ must be visited on this portion of run (acceptance-condition).


\vspace*{-10pt} 
\paragraph{Construction of $\calA_\calB^\pp$.}
Fix an extended pre-post condition $\pp=\zug{\ppre,\ppost,\final}$. Let $m_i$ be the module mapped to the $i^{\mathit{th}}$ child of the root of a strategy tree.

We construct $\calA_\calB^\pp$ such that the automaton simulates $\calB$ on an input strategy tree $T$ by using $\pp$. 
%
In particular, starting from the $i^{\mathit{th}}$ child, 
the automaton  ${\calA}_\calB^\pp$  runs in parallel a copy of $\calB$ from each control state 
$q$ such that $(m_i, q)\in \ppre$. 
When reading a node labeled with a call, $\calA_\calB^\pp$ starts at each matching return 
a copy of $\calB$ according to the applicable tuples in $\pp$ and performs updates  
according to $\final$. On all the other enabled nodes, the control state of $\calB$ is updated for each copy according to $\calB$ transitions. 


The states of $\calA_\calB^\pp$ are: an initial state $q_0$, an accepting state $q_a$, a rejecting state $q_r$, 
and states of the form $(q, \digit, f, q_{m_i},\pp)$ where
$q, q_{m_i} \in Q$, $q$ is the control state which is updated in the simulation of $\calB$, $q_{m_i}$ is the current pre-condition, and 
$\digit,f \in \{0,1\}$ are related to the winning conditions. Namely, $\digit$ is used to check the acceptance-conditions along all the plays that conform to the strategy, and  $f$ is used to expose that a final state of $\calB$ was seen between a call and its matching return. A task of $\calA_\calB^\pp$ is to handle the correct update of these bits, but they will be used to determine the acceptance by $\calA_{\calB_{win}}$.
The states $q_a$ and $q_r$ are sinks, i.e., once reached, the automaton cycles forever on them. 


\vspace*{-10pt} 
\paragraph{Construction of $\calA_{\calB,\pp}$, $\calA_{\calB,\BG}$ and $\calA_{\calB_{win}}$.}
The automaton $\calA_{\calB,\pp}$ is in charge of checking that the input tree is consistent with the extended pre-post condition $\pp$. We construct it from $\calA_\calB^\pp$ by 
modifying the transitions from a state $s$ of the form $(q, \digit, f, q_{m_i},\pp)$ at a tree-node labeled with an exit. 
In particular, in this case, we let $\calA_{\calB,\pp}$ enter $q_a$ if there exists a tuple $(m_i, q_{m_i}, ex_h, q)\in{\ppost}$, and $q_r$ otherwise.


The purpose of  $\calA_{\calB,\BG}$ is to check that the B\"uchi call 
graph $\BG$ is indeed consistent with the input strategy tree.
We construct $\calA_{\calB,\BG}$ from $\calA_\calB^\pp$ by modifying 
the transitions from calls. 
Namely,  when on a node $u$ labeled with a call to a module $m_j$ in the subtree of the $i^{th}$ child of the root, at a state of the form $(q', \digit, f, q, \pp)$ and suppose 
there is a transition $(q', \eta_{m_i}(u), q'')\in \delta$: if  
$(m_i,q)\rightarrow_F(m_j,q'')$ holds in the B\"uchi call graph $\BG$ and $\digit=0$, 
then $\calA_{\calB,\BG}$ enters the rejecting state $q_r$.
In all the other cases it behaves as $\calA_\calB^\pp$.


The purpose of $\calA_{\calB_{win}}$ is to check that the winning conditions of 
$\calB$ are satisfied along all plays of the input strategy tree. 
Again, we can modify $\calA_\calB^\pp$ to ensure this. 
In particular, when the automaton reaches a tree-node labeled with 
exit $ex$ of module $m$ in a state $(q', \digit, f, q, \pp)$, then it
enters the accepting state $q_a$, whenever $(m, q, ex, q') \in {\ppost}$ and 
 $\final(m, q, ex, q')= true$ implies $b=1$, and $q_r$ otherwise. 
Moreover the accepting states of  $\calA_{\calB_{win}}$ are $q_a$ and all the 
states of the form $(q, \digit, f, q', \pp)$ such that either $q \in F$ or $f=1$.

\vspace*{-10pt} 
\paragraph{\bf Reducing to emptiness of B\"uchi tree automata.}
$\calA_{\calB,\pp,\BG}$ can be translated into an equivalent nondeterministic 
B\"uchi tree automaton with $2^{O(|Q|^2 \log|Q|)}$ states \cite{MS95}. 
Denoting with $k$ the number of $G$ exits and  $\beta$ the number of \emph{call edges} of $G$ (i.e., the number of module pairs $(m,m')$ such that there is  
a call from $m$ to $m'$), 
the number of different choices for an extended pre-post condition 
is $2^{O(k\,|Q|^2)}\!$, for a B\"uchi call graph is $2^{2^\beta}\!$, and for a set of exits $\exits$ is $2^k\!$.  
Since $\calA_G$ and $\calA_\BG^\exits$ are both of size $O(|G|)$, the automaton $\calA_{G,\calB}$ (obtained as described earlier in this section) is of size $|G|^2\, 2^{O(|Q|^2 (k+\log |Q|)+\beta)}\!$. We can reduce the factor $|G|^2$ to $|G|$, by 
combining $\calA_G$ and $\calA_\BG^\exits$ into the same automaton (they are essentially based on $G$ transitions). Therefore, we can get an efficient construction of $\calA_{\calB,\BG}$ of size $|G|\, 2^{O(|Q|^2 (k+\log |Q|)+\beta)}$.
Thus, by Propositions \ref{prop:strat} and \ref{prop:treeStrat}, and Lemmas \ref{lemma:exits} and \ref{lemma:winning} 
\vspace*{-4pt}
\begin{theorem}\label{theo:main}
For an \RGG\ $G$ and a deterministic B\"uchi automaton $\calB$
$\pl_0$ has a winning modular strategy 
in $\zug{G,\calB}$ iff the nondeterministic B\"uchi  
tree automaton $\calA_{G,\calB}$ accepts a non-empty language. Moreover, 
$\calA_{G,\calB}$ is of size $|G|\, 2^{O(|Q|^2 (k+\log |Q|)+\beta)}$, where $k$ is the number of $G$ exits.
\end{theorem}

%

\section{Discussion}\label{sec:discussion}

In this paper, we have considered modular games with winning condition expressed by pushdown, visibly pushdown and temporal logic specifications. We have proved that the modular game problem with respect to standard pushdown specifications is undecidable. Then we have presented a number of results that give a quite accurate picture of the computational complexity of the \MVPG\ problem with visibly pushdown winning conditions. With some surprise, we have found that \MVPG\ with temporal logic winning conditions becomes immediately hard. In fact, while the complexity for \LTL\ specifications is \TWOEXPTIME-complete both for \MVPG\ and games on finite graphs, for the fragment consisting of all the Boolean combinations of \PATH-\LTL\ formulas, solving the corresponding games on finite graphs is {\sc Pspace}-complete while the \MVPG\ problem is already \TWOEXPTIME-complete. As a consequence, the computational complexity of many interesting fragments of \LTL, that have a better complexity than full \LTL\ on finite game graphs, collapses at the top of the complexities (see \cite{ltlgeneratorsJournal,concur03}).This also differs with the scenario of the complexities of model-checking RSMs in \LTL\ fragments (see \cite{icalp07}).
As a final remark, we observe that the tree automaton construction proposed in Section \ref{sec:gameToTree} can be easily adapted to handle visibly pushdown winning conditions to get a direct solution of the \MVPG\ problem.
We only need to modify the transition rules to synchronize the calls and returns of the RGG with the pushes and pops of the specification automaton, and this would be possible since they share the same visibly alphabet. The change does not affect the overall complexity, however it will slightly improve on the approach presented in Section \ref{sec:VPAreduction} that causes doubling the number of exits and gives a complexity with an exponential dependency in the number of stack symbols.

\ignore{

}

\bibliographystyle{eptcs}
\bibliography{BibGandalf}

\end{document}